\documentclass[10pt]{article}
\usepackage[latin9]{inputenc}
\pdfoutput=1  %%%% force arxiv to use pdflatex, within the first 5 lines of the preamble
\usepackage{amsmath}
\usepackage{amssymb}
\usepackage{esint}
\usepackage{subscript}

\makeatletter
\usepackage{enumitem}		% customizable list environments
\usepackage{M12}

\makeatletter
\renewcommand{\@maketitle}{% 
    \ \vspace{-3mm}
    \begin{flushleft}\LARGE%\sffamily\bfseries 
\@title \end{flushleft} 
    \vspace{9mm}
    { \addtolength{\leftskip}{25mm} \Large \noindent\ignorespaces \medskip \@author \par }
    \vspace{-3mm}
    \begin{flushright}\small \@date \mypreprintvar \end{flushright} 
    \vspace{3mm}
    \hypersetup{
    pdftitle = {\@title},
    pdfauthor = {\@author}
    }
} 
\renewcommand\LARGE{\@setfontsize\LARGE{\mytitlesize}{20pt}} %% here 20 is line spacing 
\renewcommand\Large{\@setfontsize\Large{\myauthorsize}{16pt}} 
\makeatother

\titleformat*{\section}{\Large\mdseries\itshape}

\makeatletter
\newsavebox\myboxA
\newsavebox\myboxB
\newlength\mylenA

\newcommand*\xoverline[2][0.75]{%
    \sbox{\myboxA}{$\m@th#2$}%
    \setbox\myboxB\null% Phantom box
    \ht\myboxB=\ht\myboxA%
    \dp\myboxB=\dp\myboxA%
    \wd\myboxB=#1\wd\myboxA% Scale phantom
    \sbox\myboxB{$\m@th\overline{\copy\myboxB}$}%  Overlined phantom
    \setlength\mylenA{\the\wd\myboxA}%   calc width diff
    \addtolength\mylenA{-\the\wd\myboxB}%
    \ifdim\wd\myboxB<\wd\myboxA%
       \rlap{\hskip 0.5\mylenA\usebox\myboxB}{\usebox\myboxA}%
    \else
        \hskip -0.5\mylenA\rlap{\usebox\myboxA}{\hskip 0.5\mylenA\usebox\myboxB}%
    \fi}
\makeatother

\DeclareFontFamily{U}{mathx}{\hyphenchar\font45}
\DeclareFontShape{U}{mathx}{m}{n}{
      <5> <6> <7> <8> <9> <10>
      <10.95> <12> <14.4> <17.28> <20.74> <24.88>
      mathx10
      }{}
\DeclareSymbolFont{mathx}{U}{mathx}{m}{n}
\DeclareFontSubstitution{U}{mathx}{m}{n}
\DeclareMathAccent{\widebar}{0}{mathx}{"73}

\AtBeginDocument{
  
}

\makeatother

\begin{document}
\settitlesize{17pt}

\title{\textrm{An improved AFS phase for AdS\textsubscript{3} string integrability}\vspace{-3mm}}

\author{Michael C. Abbott\alabel{$\,1$} \emph{\&} \  In\^{e}s Aniceto\alabel{$\,2$}
 \address[$1$]{Department of Mathematics, University of Cape Town,\\
Rondebosch 7701, Cape Town, South Africa. } \address[$2$]{Institute
of Physics, Jagiellonian University, \\
Ul. \L{}ojasiewicza 11, 30-348 Krak\'{o}w, Poland.}
%\address{michael.abbott@uct.ac.za, ianiceto@math.ist.ut}
 }

\date{22 December 2014}
\maketitle
\begin{abstract}
We propose a number of modifications to the classical term in the
dressing phase for integrable strings in $AdS_{3}\times S^{3}\times S^{3}\times S^{1}$,
and check these against existing perturbative calculations, crossing
symmetry, and the semiclassical limit of the Bethe equations. The
principal change is that the phase for different masses should start
with a term $Q_{1}Q_{2}$, like the one-loop $AdS_{3}$ dressing phase,
rather than $Q_{2}Q_{3}$ as for the original $AdS_{5}$ AFS phase. 
\end{abstract}
\noindent The central object in the integrable picture of planar
AdS$_{5}$/CFT$_{4}$ is the all-loop S-matrix, and the Bethe ansatz
equations which follow from this \cite{Minahan:2002ve,Beisert:2010jr}.
Its nontrivial dependence on the 't Hooft coupling $\lambda$ comes
from the dressing phase, and expanding at strong coupling this has
the form 
\begin{equation}
\sigma_{\text{BES}}(x,y)=\exp\left[i\frac{\sqrt{\lambda}}{2\pi}\smash{\sum_{r,s\geq2}}c_{r,s}(\lambda)\: Q_{r}(x)Q_{s}(y)\right]\label{eq:sigma-AdS5}
\end{equation}
where $c_{r,s}=\left(\delta_{r+1,s}-\delta_{r,s+1}\right)+a_{r,s}/\sqrt{\lambda}+\bigo{1/\lambda}$.
The first term was introduced by Arutyunov, Frolov and Staudacher
(AFS) in \cite{Arutyunov:2004vx} as a correction needed to match
classical strings in $AdS_{5}\times S^{5}$. The coefficients $a_{r,s}$
are the extension to one-loop strings of \cite{Beisert:2005cw,Hernandez:2006tk},
and this was later extended to all loops in \cite{Beisert:2006ez,Beisert:2006ib}.
%\vphantom{\cite{Babichenko:2009dk, Borsato:2012ud, Borsato:2012ss}\cite{Abbott:2013mpa}\cite{Borsato:2013qpa, Borsato:2013hoa, Sfondrini:2014via}}
% this is just to make these show up together, in order!

The dressing phase for $AdS_{3}$ backgrounds is different, and is
now understood quite well at one loop \cite{Borsato:2013hoa,Abbott:2013mpa,Sundin:2014sfa,Bianchi:2014rfa};
see also \cite{David:2010yg,Beccaria:2012kb}. However we believe
that the classical part of the dressing phase has been treated incorrectly
in the literature. This is the subject of our letter. 

A new feature of strings in $AdS_{3}\times S^{3}\times S^{3}\times S^{1}$
is that there are excitations (above the BMN state) of mass $1$,
$\alpha$, $1-\alpha$ and $0$ \cite{Babichenko:2009dk}, rather
than just one mass in $AdS_{5}\times S^{5}$ or two in $AdS_{4}\times CP^{3}$.
The bosonic modes of mass $s_{1}=\alpha$ and $s_{3}=1-\alpha$ are
excitations in the two $S^{3}$ factors (which have different radii),
and there are two such excitations in each sphere, one in the left
copy of the algebra (labelled $1$, or $3$) and one in the right
($\bar{1}$, or $\bar{3}$). These and their superpartners are the
elementary particles in the Bethe ansatz description of \cite{Borsato:2012ss},
which gives the spectrum as 
\begin{equation}
\Delta-J=\sum_{\ell}\sum_{k=1}^{K_{\ell}}E_{\ell}(p_{\ell,k}),\qquad E_{\ell}(p_{\ell,k})=\sqrt{s_{\ell}^{2}+4h^{2}\sin^{2}\frac{p_{\ell,k}}{2}}
\label{eq:Disp-rel}
\end{equation}
where the allowed $p_{\ell,k}$ are constrained by equations of the
form $e^{ip_{\ell,k}L}=\prod_{j\neq k}S(p_{k},p_{j})$, using the
S-matrix of \cite{Borsato:2012ud}. This must include (for the first
time%
\footnote{In the $AdS_{4}\times CP^{3}$/ABJM correspondence there are particles
of mass $1$ and $\tfrac{1}{2}$, but only the latter appear in the
Bethe equations, and hence in the AFS phase. The heavy particles are
composite objects, mirror bound states \cite{Zarembo:2009au,Abbott:2011tp}.
The entire dressing phase for this correspondence is simply half the
BES phase \cite{Gromov:2008qe,Mauri:2013vd}. %
}) a dressing phase for the scattering of particles of different mass.

\newcommand{\cJ}{\mathcal{J}}
\newcommand{\cS}{\mathcal{S}}
\newcommand{\blmt}{\text{\tiny{BLMT}}}
\newcommand{\bosst}{\text{\scriptsize{BOSST}}}

\newcommand{\Gsuper}[1]{{G^{(#1)}}} 
\newcommand{\ghl}{{g_\text{\tiny{HL}}}}

\newcommand{\Aws}{A_\text{WS}}

\newcommand{\xhl}{\tilde{x}}
\newcommand{\qhl}{\tilde{q}}
\newcommand{\pbmn}{\tilde{p}}

\newcommand{\oneover}[1]{\scalebox{0.85}{\raisebox{1mm}{1}\hspace{-0.5mm}\big/$#1$}}

The first classical phase for two particles of different mass was
written down by Borsato, Ohlsson Sax and Sfondrini \cite{Borsato:2012ss},
who gave 
\begin{equation}
\sigma_{\text{BOS}}(x,y)=\left(\frac{1-\frac{1}{x^{+}y^{-}}}{1-\frac{1}{x^{+}y^{+}}}\:\frac{1-\frac{1}{x^{-}y^{+}}}{1-\frac{1}{x^{-}y^{-}}}\right)^{i\frac{h}{W_{xy}}\big(x+\frac{1}{x}-y-\frac{1}{y}\big)}\left(\frac{1-\frac{1}{x^{-}y^{+}}}{1-\frac{1}{x^{+}y^{-}}}\right)\label{eq:sigma-BOS}
\end{equation}
where the masses $s_{x}$, $s_{y}$ enter explicitly through 
\[
W_{xy}=\frac{4s_{x}s_{y}}{s_{x}+s_{y}}=\begin{cases}
2s_{x}, & s_{x}=s_{y}\\
4s_{x}s_{y}, & s_{x}+s_{y}=1.
\end{cases}
\]
When $s_{x}=s_{y}=1$ this is exactly the original AFS phase used
in $AdS_{5}$. A similar phase was used by \cite{Sundin:2013ypa}
when comparing to tree-level amplitudes, but with the exponent of
the first factor replaced by%
\footnote{The variables $x^{\pm}$ depend on the mass $s_{x}$ through 
\[
x^{\pm}+\frac{1}{x^{\pm}}=x+\frac{1}{x}\pm i\frac{s_{x}}{h}
\]
where $h=\sqrt{\lambda}/2\pi+c+\bigo{1/\sqrt{\lambda}}$ is the Bethe
coupling, normalised as in \cite{Borsato:2012ud,Abbott:2012dd,Borsato:2012ss,Sundin:2013ypa}. %
} 
\[
\frac{ih}{W_{xy}}\Big(x^{+}+\frac{1}{x^{+}}-y^{-}-\frac{1}{y^{-}}\Big)=\frac{ih}{W_{xy}}\Big(x+\frac{1}{x}-y-\frac{1}{y}\Big)-\frac{s_{x}+s_{y}}{W_{xy}}.
\]
The last term here has no effect on tree-level worldsheet scattering. 

Our first proposal is that the correct generalisation of the AFS phase
to particles of different mass is instead: 
\begin{equation}
\sigma_{\text{AFS}}(x,y)=\left(\frac{1-\frac{1}{x^{+}y^{-}}}{1-\frac{1}{x^{+}y^{+}}}\:\frac{1-\frac{1}{x^{-}y^{+}}}{1-\frac{1}{x^{-}y^{-}}}\right)^{i\frac{h}{W_{xy}}\big(x+\frac{1}{x}-y-\frac{1}{y}\big)}\left(\frac{1-\frac{1}{x^{+}y^{+}}}{1-\frac{1}{x^{-}y^{-}}}\right)^{\negthickspace\frac{s_{x}-s_{y}}{W_{xy}}}\left(\frac{1-\frac{1}{x^{-}y^{+}}}{1-\frac{1}{x^{+}y^{-}}}\right)^{\negthickspace\frac{s_{x}+s_{y}}{W_{xy}}\negthickspace\negthickspace}.\label{eq:our-AFS-phase}
\end{equation}
This follows from changing the original definition, the first term of \eqref{eq:sigma-AdS5}, by an overall
factor: 

\begin{equation}
\sigma_{\text{AFS}}(x,y)=\exp\Big\{ i\frac{h}{W_{xy}}\sum_{r=2}^{\infty}\left[Q_{r}(x)Q_{r+1}(y)-Q_{r+1}(x)Q_{r}(y)\right]\Big\}.\label{eq:AFS-sum-QQ}
\end{equation}
However this change alone will break the agreement with tree-level
worldsheet scattering seen in \cite{Sundin:2013ypa}, as we discuss
below. This leads us to suggest two further modifications, which we
parameterise by $\beta,\delta,\Delta$, in addition to \cite{Borsato:2012ss}'s
$\gamma,\Gamma$. Of these five parameters, three will be fixed by
tree-level scattering, and one more by a semiclassical limit of the
Bethe equations. 
\begin{itemize}
\item In the one-loop dressing phase, an important difference from the $AdS_{5}$
case is that the sum starts with $a_{1,2}Q_{1}Q_{2}$ \cite{Beccaria:2012kb,Borsato:2013hoa,Abbott:2013mpa},
rather than $a_{2,3}Q_{2}Q_{3}$ as in \eqref{eq:sigma-AdS5}. It
seems natural to wonder if this should apply to the classical phase
too, and thus our second proposal is to include a factor
\begin{equation}
\sigma_{\text{one}}(x,y)=\exp\Big\{ i\frac{h}{W_{xy}}\left[p_{x}\: Q_{2}(y)-p_{y}\: Q_{2}(x)\right]\Big\}.\label{eq:our-beta-phase}
\end{equation}
We use $\sigma_{\text{one}}^{\,\beta}\:\sigma_{\text{AFS}}$ as the
classical phase for different-mass scattering only, with power $\beta=1$
most natural. 
\item The S-matrix derived by \cite{Borsato:2012ud} contains a number of
unfixed scalars $S^{\ell m}$, each of which should include the dressing
phase. An ansatz for the remaining factors was given by \cite{Borsato:2012ss},
and our third proposal is that this should be slightly modified, introducing
a phase like the one needed for the string frame, but with an arbitrary
power. Explicitly, we set 
\begin{equation}
\begin{aligned}S^{11}(x,y) & =\left(\frac{x^{-}y^{+}}{x^{+}y^{-}}\right)^{\frac{1}{2}+\gamma+\delta}\left[\frac{1-\oneover{x^{+}y^{-}}}{1-\oneover{x^{-}y^{+}}}\:\sigma_{\text{AFS}}^{\,2}(x,y)\right]^{1+2\gamma}\sigma_{LL}^{\,2}(x,y)\\
S^{13}(x,y) & =\left(\frac{x^{-}y^{+}}{x^{+}y^{-}}\right)^{\Gamma+\Delta}\left[\frac{1-\oneover{x^{+}y^{-}}}{1-\oneover{x^{-}y^{+}}}\:\sigma_{\text{one}}^{\,2\beta}(x,y)\:\sigma_{\text{AFS}}^{\,2}(x,y)\right]^{1+2\Gamma}\sigma_{LL}^{\,2}(x,y).
\end{aligned}
\label{eq:our-S11-ansatz}
\end{equation}
The expressions in \cite{Borsato:2012ss} have unfixed $\gamma$ and
$\Gamma$ but $\delta=\Delta=0$, while going to the string frame
would normally mean increasing $\delta$ and $\Delta$ by $\tfrac{1}{2}$.
We write the one-loop dressing phase $\sigma_{LL}$ outside the power
of $1+2\gamma$, as it was the total phase which was calculated by
semiclassical means in \cite{Abbott:2013mpa}. We omit the two-loop
and higher phases.
\end{itemize}

\section*{Tree-level BMN Scattering}

Let us now test this against the results of Sundin and Wulff \cite{Sundin:2013ypa},
who computed tree-level Feynman diagrams in the worldsheet theory.
To do this we must take the BMN limit, writing $p=\pbmn/h$ with $\pbmn$
order $1$ and $h\gg1$. Then we can expand 
\[
x^{\pm}=\frac{s_{x}+\omega_{x}}{\pbmn_{x}}\pm\frac{i(s_{x}+\omega_{x})}{2h}+\bigodiv{h^{2}},\qquad\mbox{where}\quad\omega_{x}\equiv\sqrt{s_{x}^{2}+\pbmn_{x}^{2}}=E_{x}(p_{x})+\ldots.
\]
The charges used above are $Q_{1}(x)\equiv p_{x}=-i\log(\xp/\xm)$
and, for $n>1$, 
\[
Q_{n}(x)\equiv\frac{i}{n-1}\left[\frac{1}{(\xp)^{n-1}}-\frac{1}{(\xm)^{n-1}}\right]=\frac{\pbmn_{x}}{h}\Big(\frac{\omega_{x}-s_{x}}{\pbmn_{x}}\Big)^{n-1}+\frac{0}{h^{2}}+\bigodiv{h^{3}}.
\]
Apart from obvious phases, the other expansions we will need for this
limit are 
\begin{align*}
\negthickspace\negthickspace\frac{1-\oneover{x^{+}y^{-}}}{1-\oneover{x^{-}y^{+}}}\left[\sigma_{\text{one}}^{\,\beta}\:\sigma_{\text{AFS}}\right]^{2} & =1+\frac{i}{2h}\left[-\pbmn_{x}(\omega_{y}-s_{y})\Big(\frac{1}{s_{x}}-\frac{4\beta}{W_{xy}}\Big)+\pbmn_{y}(\omega_{x}-s_{x})\Big(\frac{1}{s_{y}}-\frac{4\beta}{W_{xy}}\Big)\right]+\ldots\displaybreak[0]\\
\frac{x^{+}-y^{-}}{x^{-}-y^{+}} & =1+\frac{i}{2h}\left[\pbmn_{x}-\pbmn_{y}+\frac{\alpha(\pbmn_{x}+\pbmn_{y})^{2}}{\omega_{x}\pbmn_{y}-\omega_{y}\pbmn_{x}}\right]+\bigodiv{h^{2}},\qquad s_{x}=s_{y}=\alpha\;\mbox{only}.
\end{align*}

Consider two bosons from the left sector of the theory, ``1'' of
mass $\alpha$ and ``3'' of mass $1-\alpha$. As in \cite{Sundin:2013ypa},
and in \cite{Arutyunov:2009ga,Zarembo:2009au}, we should allow for
some unknown gauge dependence through $\tilde{a}$ in addition to
the spin-chain S-matrix. However for the mixed-mass case we allow
two parameters $\smash{\tilde{b}},\tilde{c}$ (and expect them to
be equal at $\alpha=\tfrac{1}{2}$). Thus we write the scattering
amplitudes as 
\begin{equation}
\begin{aligned}A^{11}(x,y) & =\exp\Big[-\frac{i\tilde{a}}{h\alpha}\big(\omega_{x}\pbmn_{y}-\omega_{y}\pbmn_{x}\big)\Big]\:\frac{x^{+}-y^{-}}{x^{-}-y^{+}}\: S^{11}(x,y), &  & s_{x}=s_{y}=\alpha\\
A^{13}(x,y) & =\exp\Big[-\frac{i}{h}\Big(\tilde{c}\frac{\omega_{x}\pbmn_{y}}{\alpha}-\tilde{b}\frac{\omega_{y}\pbmn_{x}}{1-\alpha}\Big)\Big]\: S^{13}(x,y), &  & s_{x}=\alpha,\; s_{y}=1-\alpha.
\end{aligned}
\label{eq:Amp-with-Shat+atil}
\end{equation}
The corresponding worldsheet results are in equation (3.2) of \cite{Sundin:2013ypa}.
These depend on the AFZ gauge parameter \cite{Arutyunov:2006gs} which
is $a=\frac{1}{2}$ for the simplest light-cone gauge:%
\footnote{These are $A^{(22)}$ and $A^{(23)}$ in the notation of \cite{Sundin:2013ypa},
where the particle of mass $\alpha$ is ``2''. We have also restored
a factor $1/h$. %
}
\begin{align*}
\Aws^{11}(\pbmn_{x},\pbmn_{y}) & =1+\frac{i}{2h}\frac{\alpha(\pbmn_{x}+\pbmn_{y})^{2}}{\omega_{x}\pbmn_{y}-\omega_{y}\pbmn_{x}}+\frac{i}{2h}(1-2a)\left[\omega_{x}\pbmn_{y}-\omega_{y}\pbmn_{x}\right]+\bigodiv{h^{2}}\displaybreak[0]\\
\Aws^{13}(\pbmn_{x},\pbmn_{y}) & =1+\frac{i}{2h}(1-2a)\left[\omega_{x}\pbmn_{y}-\omega_{y}\pbmn_{x}\right]+\bigodiv{h^{2}}.
\end{align*}
Matching $A^{11}=\Aws^{11}$ and $A^{13}=\Aws^{13}$, and demanding
that $\Gamma\neq-\frac{1}{2}$, we find that 
\begin{equation}
\beta=1,\qquad\delta_{\text{SF}}=\frac{1}{2}\qquad\Delta_{\text{SF}}=-\frac{1}{2}-2\Gamma\vspace{-5mm}\label{eq:our-beta-gamma-results}
\end{equation}
and 
\[
2\tilde{a}=1+2\gamma+(2a-1)\alpha,\quad2\tilde{b}=-(1+2\Gamma)+(2a-1)(1-\alpha),\quad2\tilde{c}=-(1+2\Gamma)+(2a-1)\alpha.
\]
We write $\delta_{\text{SF}}$ to indicate that these are the parameters
in the string frame; in the spin chain frame we have $\delta=0$ and
$\Delta=-1-2\Gamma$ instead. 

The comparison performed by \cite{Sundin:2013ypa} used $\sigma_{\text{BOS}}$
for $A^{13}$ (and for $A^{11}$, $\sigma_{\text{BOS}}=\sigma_{\text{AFS}}$).
If we repeat this allowing arbitrary parameters (including $\beta$,
and demanding $\alpha\neq\tfrac{1}{2}$, $\Gamma\neq-\tfrac{1}{2}$)
we find that 
\[
\beta=0,\qquad\delta_{\text{SF}}=\Delta_{\text{SF}}=\tfrac{1}{2}
\]
and $2\tilde{a}=1+2\gamma+(2a-1)\alpha$, $2\tilde{b}=1+2\Gamma+(2a-1)(1-\alpha)$,
$2\tilde{c}=1+2\Gamma+(2a-1)\alpha$. Setting $\gamma=\Gamma=0$ returns
precisely the phases used by \cite{Sundin:2013ypa}. 

We can similarly check agreement for scattering with a ``$\bar{1}$''
or ``$\bar{3}$'' particle in the right sector, using the same gauge
phases $\tilde{a}$, $\tilde{b}$, $\tilde{c}$ as before with the
appropriate $\hat{S}$ matrix elements from \cite{Borsato:2012ud}:
\begin{align*}
A^{1\bar{1}}(x,y) & =e^{-\frac{i\tilde{a}}{h\alpha}(\omega_{x}\pbmn_{y}-\omega_{y}\pbmn_{x})}\:\frac{\sqrt{1-\oneover{x^{+}y^{+}}}\sqrt{1-\oneover{x^{-}y^{-}}}}{1-\oneover{x^{+}y^{-}}}\: S^{1\bar{1}}(x,y)\displaybreak[0]\\
A^{1\bar{3}}(x,y) & =e^{-\frac{i}{h}(\tilde{c}\frac{\omega_{x}\pbmn_{y}}{\alpha}-\tilde{b}\frac{\omega_{y}\pbmn_{x}}{1-\alpha})}\:\frac{\sqrt{1-\oneover{x^{+}y^{+}}}\sqrt{1-\oneover{x^{-}y^{-}}}}{1-\oneover{x^{+}y^{-}}}\: S^{1\bar{3}}(x,y).
\end{align*}
The phases $S^{\ell\bar{m}}$ should be modified from those of \cite{Borsato:2012ss}
by the same factors $\delta,\Delta$, i.e. 
\begin{equation}
S^{1\bar{1}}(x,y)=\left[\frac{1-\oneover{x^{+}y^{-}}}{1-\oneover{x^{-}y^{+}}}\right]^{-\frac{1}{2}}\negthickspace S^{11}(x,y),\qquad S^{1\bar{3}}(x,y)=\left[\frac{1-\oneover{x^{+}y^{-}}}{1-\oneover{x^{-}y^{+}}}\right]^{+\frac{1}{2}}\negthickspace S^{13}(x,y)\label{eq:S11bar-etc}
\end{equation}
(and $\sigma_{LL}$ is replaced with $\sigma_{LR}$) and the worldsheet
results are \cite{Sundin:2013ypa} 
\[
\Aws^{1\bar{1}}(\pbmn_{x},\pbmn_{y})=\Aws^{11}(\pbmn_{x},\pbmn_{y})-\frac{i}{2h}\frac{4\alpha\pbmn_{x}\pbmn_{y}}{\omega_{x}\pbmn_{y}-\omega_{y}\pbmn_{x}}+\bigodiv{h},\qquad\Aws^{1\bar{3}}(\pbmn_{x},\pbmn_{y})=\Aws^{13}(\pbmn_{x},\pbmn_{y}).
\]
Clearly we obtain no new constraints from these.

\section*{Crossing Relations}

We can obtain a check on the phases described above from crossing
symmetry \cite{Janik:2006dc,Arutyunov:2006iu,Vieira:2010kb}. If we
stay in the BMN limit there is nothing to learn, since (by construction)
we have not changed the results. But if we take the semiclassical
limit without small momentum ($h\gg1$, $p\sim1$) then we obtain
a nontrivial check which in fact mixes the classical and one-loop
phases. The relevant equations from \cite{Borsato:2012ss} for the
scalars $S^{\ell m}$ \eqref{eq:our-S11-ansatz} and $S^{\ell\bar{m}}$
\eqref{eq:S11bar-etc} are 
\begin{equation}
\negthickspace\negthickspace\negthickspace\begin{aligned}S^{11}(x,y)S^{1\bar{1}}(x,\bar{y}) & =\frac{x^{-}-y^{+}}{x^{-}-y^{-}}\sqrt{\frac{x^{+}}{x^{-}}}\sqrt{\frac{x^{-}-y^{-}}{x^{+}-y^{+}}} &  & \negthickspace=i\, e^{i(p_{x}-p_{y})/4}\frac{1-e^{i(p_{x}+p_{y})/2}}{1-e^{i(p_{x}-p_{y})/2}}+\bigodiv{h}\\
S^{13}(x,y)S^{1\bar{3}}(x,\bar{y}) & =\frac{x^{+}-y^{-}}{x^{-}-y^{-}}\sqrt{\frac{x^{+}}{x^{-}}}\sqrt{\frac{x^{-}-y^{-}}{x^{+}-y^{+}}} &  & \negthickspace=-i\, e^{i(3p_{x}-3p_{y})/4}\frac{1-e^{i(p_{x}+p_{y})/2}}{1-e^{i(p_{x}-p_{y})/2}}+\ldots.
\end{aligned}
\negthickspace\negthickspace\negthickspace\negthickspace\label{eq:crossing-equations}
\end{equation}
Here $\bar{y}$ indicates that the argument has been moved $y^{\pm}\to1/y^{\pm}$.
On the right we use $x^{\pm}=e^{\pm ip_{x}/2}+\bigo{1/h}$, and separate
two factors: a phase and a trigonometric part. (There are two more
crossing equations, for $S^{11}(x,\bar{y})S^{1\bar{1}}(x,y)$ and
$S^{13}(x,\bar{y})S^{1\bar{3}}(x,y)$. These can be treated almost
identically.)

In this $h\gg1$ limit we can write the complete dressing phase as
\[
\sigma_{\text{one}}^{\,\beta}\:\sigma_{\text{AFS}}\:\sigma_{LL}\:\sigma_{\text{higher-loop}}=\exp\left[ih(\beta\theta_{\text{one}}+\theta_{\text{AFS}})+i\theta_{LL}+\bigo{1/h}\vphantom{1^{1}}\right]
\]
with each $\theta$ of order 1. Considering \eqref{eq:crossing-equations}
at order $h$ in the exponent, the cancellation is very simple from
\eqref{eq:AFS-sum-QQ} and \eqref{eq:our-beta-phase}, because $Q_{n}(1/y^{\pm})=-Q_{n}(y^{\pm})+\bigo{1/h}$.
At order $h^{0}$ it's easier to use form \eqref{eq:our-AFS-phase}
for the AFS phase. The exponent $\frac{ih}{W_{xy}}(x+\frac{1}{x}-y-\frac{1}{y})$
has terms at order $h$ and $h^{-1}$ but not $h^{0}$, so this first
factor does not contribute. The other two factors give 
\begin{align*}
\sigma_{\text{AFS}}(x^{\pm},y^{\pm})\times\sigma_{\text{AFS}}\Big(x^{\pm},\frac{1}{y^{\pm}}\Big) & =\left(\frac{1-\frac{1}{x^{+}y^{+}}}{1-\frac{1}{x^{-}y^{-}}}\times\frac{1-\frac{y^{+}}{x^{+}}}{1-\frac{y^{-}}{x^{-}}}\right)^{\negthickspace\frac{s_{x}-s_{y}}{W_{xy}}}\left(\frac{1-\frac{1}{x^{-}y^{+}}}{1-\frac{1}{x^{+}y^{-}}}\times\frac{1-\frac{y^{+}}{x^{+}}}{1-\frac{y^{-}}{x^{-}}}\right)^{\negthickspace\frac{s_{x}+s_{y}}{W_{xy}}}\\
 & =\exp\left[i\frac{2p_{x}s_{y}}{W_{xy}}+\bigodiv{h}\right].
\end{align*}
At the same order there is also a contribution from \eqref{eq:our-beta-phase}.
Using $Q_{2}(1/y^{\pm})=-Q_{2}(y^{\pm})-2s_{y}/h+\bigo{1/h^{2}}$
we see that it exactly cancels the last equation if $\beta=1$: 
\begin{equation}
\sigma_{\text{one}}(x^{\pm},y^{\pm})\:\sigma_{\text{one}}\Big(x^{\pm},\frac{1}{y^{\pm}}\Big)=\exp\Big(-i\frac{2p_{x}s_{y}}{W_{xy}}+\ldots\Big).\label{eq:crossing-beta-contrib}
\end{equation}
Note that if $\beta=0$, it is difficult to imagine what would cancel
the phase $e^{ip_{x}/2\alpha}$ from $\sigma_{\text{AFS}}$ in the
$S^{13}S^{1\bar{3}}$ case at generic $\alpha$.%
\footnote{If we used \eqref{eq:sigma-BOS} instead, the power would be an integer:
$\sigma_{\text{BOS}}(x^{\pm},y^{\pm})\:\sigma_{\text{BOS}}\big(x^{\pm},\tfrac{1}{y^{\pm}}\big)=e^{ip_{x}}+\bigo{1/h}$
in both the $S^{11}S^{1\bar{1}}$ and $S^{13}S^{1\bar{3}}$ cases.%
} 

For the remaining factors in $S^{\ell m}$ \eqref{eq:our-S11-ansatz}
and $S^{\ell\bar{m}}$ \eqref{eq:S11bar-etc}, the contribution is
\begin{align*}
S^{11}S^{1\bar{1}}:\qquad & \left(\frac{x^{-}y^{+}}{x^{+}y^{-}}\right)^{\frac{1}{2}+\gamma+\delta}\left[\frac{1-\frac{1}{x^{+}y^{-}}}{1-\frac{1}{x^{-}y^{+}}}\right]^{1+2\gamma}\left(\frac{x^{-}y^{-}}{x^{+}y^{+}}\right)^{\frac{1}{2}+\gamma+\delta}\left[\frac{1-\frac{y^{-}}{x^{+}}}{1-\frac{y^{+}}{x^{-}}}\right]^{-\frac{1}{2}+2\gamma}\\
 & =i\:\exp\left[-ip_{x}(7/4+4\gamma+2\delta)+ip_{y}/4\right]+\bigo{1/h}.
\end{align*}
Combined with $(e^{ip_{x}})^{2(1+2\gamma)}$ from $\sigma_{\text{AFS }}$,
and using coefficients \eqref{eq:our-beta-gamma-results} with the
spin-chain-frame $\delta=0$, we get $e^{ip_{x}/4}$ as in \eqref{eq:crossing-equations}.
For the mixed mass case, the remaining contribution is instead 
\[
S^{13}S^{1\bar{3}}:\qquad-i\:\exp\left[-ip_{x}(5/4+4\Gamma+2\Delta)-ip_{y}/4\right]+\bigo{1/h}
\]
which combined with $\sigma_{\text{one}}\sigma_{\text{AFS }}$ gives
$e^{i3p_{x}/4}$. In both cases the power of $e^{ip_{y}}$ does not
yet match \eqref{eq:crossing-equations}.

At order $h^{0}$ there will also be a contribution from the one-loop
phase. The semiclassical calculation of this in \cite{Abbott:2013mpa}
gave the following final answer for left-left scattering:\newcommand{\Itil}{\widetilde{I}}
\begin{equation}
\begin{aligned}\theta_{LL}(x^{\pm},y^{\pm}) & =\chi(x^{+},y^{+})-\chi(x^{+},y^{-})-\chi(x^{-},y^{+})+\chi(x^{-},y^{-})\\
 & =\left(I_{yx}-I_{xy}\right),\qquad I_{yx}=\sum_{\pm}\frac{\mp1}{16\pi}\int_{U_{\pm}}\negthickspace dz\:\frac{\partial G(z,y^{\pm})}{\partial z}\: G(z,x^{\pm})
\end{aligned}
\label{eq:theta_HL-my-integrals}
\end{equation}
and for left-right scattering: 
\begin{align*}
\theta_{LR}(x^{\pm},y^{\pm}) & =\tilde{\chi}(x^{+},y^{+})-\tilde{\chi}(x^{+},y^{-})-\tilde{\chi}(x^{-},y^{+})+\tilde{\chi}(x^{-},y^{-})\\
 & =\left(\Itil_{yx}-\Itil_{xy}\right),\qquad\Itil_{yx}=\sum_{\pm}\frac{\mp1}{16\pi}\int_{U_{\pm}}\negthickspace dz\:\frac{\partial G(z,y^{\pm})}{\partial z}\: G\Big(\frac{1}{z},x^{\pm}\Big)
\end{align*}
where 
\[
G(z,x^{\pm})\equiv-i\log\left(\frac{z-x^{+}}{z-x^{-}}\right)-\frac{p_{x}}{2}.
\]
Notice that $G(\tfrac{1}{z},x^{\pm})=G(z,1/x^{\pm})$. Then it is
easy to see that $\theta_{LL}(x^{\pm},y^{\pm})+\theta_{LR}(x^{\pm},1/y^{\pm})=0$,
and thus there is no contribution to crossing from evaluating at $1/y^{\pm}$.
However in moving $y^{\pm}\to1/y^{\pm}$ we move some poles across
contours.  

Let us focus on the effect on the term $\tilde{\chi}(x^{+},y^{+})$.
The only pole in the integrand at $z=y^{+}$ comes from $\partial_{z}G(z,y^{\pm})$
in $\Itil_{yx}$. Moving the pole to $z=1/y^{+}$ pulls it across
$U_{+}$ anti-clockwise, and the final pole has residue $-iG(y^{+},x^{\pm})$.
The contribution is then 
\[
\Delta\tilde{\chi}(x^{+},y^{+})=\frac{i}{8}\:\left[-\log(y^{+}-x^{+})+\frac{1}{2}\log x^{+}\right]
\]
 There is a similar contribution from $\Itil_{xy}$, from the log
cut.  Together these give the remainder of \eqref{eq:crossing-equations}:
\begin{equation}
\sigma_{LL}^{\,2}(x,y)\sigma_{LR}^{\,2}(x,\bar{y})=\frac{\sqrt{x^{+}-y^{-}}\sqrt{x^{-}-y^{+}}}{\sqrt{x^{+}-y^{+}}\sqrt{x^{-}-y^{-}}}=e^{-ip_{y}/2}\frac{1-e^{i(p_{x}+p_{y})/2}}{1-e^{i(p_{x}-p_{y})/2}}\label{eq:crossing-trig}
\end{equation}

\vspace{3pt}

\section*{Semiclassical Limit of Bethe Equations}

Another check of the phases is to look at the semiclassical limit
of the Bethe equations, which should reproduce the finite-gap equations.
This calculation was also done by \cite{Borsato:2012ss}, so we do
not show much detail. But the result is changed by using our phase:
\cite{Borsato:2012ss} found $\Gamma=\gamma+\tfrac{1}{2}$. 

It suffices to look at the left sector, with $K_{1}\neq0$ and $K_{3}\neq0$
only. Then equations (4.5) and (4.7) of \cite{Borsato:2012ss} become
\begin{align}
\frac{2\pi n_{1,k}}{2\alpha} & =\frac{-x}{x^{2}-1}\Big\{\left[L+K_{1}(\tfrac{1}{2}+\gamma+\delta)+K_{3}(\Gamma+\Delta)\right]+Q_{1,2}\left[1+(1+2\gamma)\right]\nonumber \\
 & \qquad+Q_{3,2}\left[(1+2\Gamma)\frac{1-\alpha-\beta}{\alpha}\right]\Big\}+\frac{-1}{x^{2}-1}\frac{(1+2\Gamma)}{\alpha}\Big[\alpha Q_{1,1}+(\beta-\alpha)Q_{3,1}\Big]\nonumber \\
 & \qquad+2\fint dy\frac{\rho_{1}(y)}{x-y}-\frac{(1+\Gamma)}{\alpha}\Big[\alpha Q_{1,1}+(1-\alpha)Q_{3,1}\Big]\label{eq:semiclass-BAE-first}\\
\frac{2\pi n_{3,k}}{2(1-\alpha)} & =\frac{-x}{x^{2}-1}\Big\{\left[L+K_{1}(\Gamma+\Delta)+K_{3}(\tfrac{1}{2}+\gamma+\delta)\right]+Q_{3,2}\left[1+(1+2\gamma)\right]\nonumber \\
 & \qquad+Q_{1,2}\left[(1+2\Gamma)\frac{\alpha-\beta}{1-\alpha}\right]\Big\}+\frac{1}{x^{2}-1}\left[\mbox{winding}\right]+2\fint dy\frac{\rho_{3}(y)}{x-y}+\left[\mbox{constant}\right]\nonumber 
\end{align}
where $Q_{\ell,n}$ is the total charge $Q_{n}$ of particles of type
$\ell$ (and of course $Q_{1}$ is momentum, $Q_{2}$ an energy).
Define $\mathcal{E}_{\ell}$ to be the curly brackets above (i.e.
$-\tfrac{1}{2}$ the sum of the residues at $x=\pm1$, divided by
the mass). 

If we set $\mathcal{E}_{1}=\mathcal{E}_{3}$ (which in the language
of \cite{Abbott-2014-ip-2} means working above the $\zeta=\phi$
vacuum) we find 
\begin{equation}
\beta=1,\qquad\gamma+\Gamma=-\frac{3}{2},\qquad\delta-\Delta=1+2\Gamma.\label{eq:semiclass-BAE-gamma-etc}
\end{equation}
We have derived these constraints on the parameters independent of
the near-BMN comparison, \eqref{eq:our-beta-gamma-results}, but the
two are clearly compatible. Using both (i.e. using \eqref{eq:semiclass-BAE-gamma-etc}
and $\delta=0$) we get
\[
2\pi\mathcal{E}_{1}=2\pi\mathcal{E}_{3}=L-(1+\Gamma)\left(K_{1}+K_{3}\right)-(1+2\Gamma)\left(Q_{1,2}+Q_{3,2}\right).
\]

\vspace{3pt}

\section*{Conclusion}

In summary, we suggest three alterations to the classical dressing
phase given by \cite{Borsato:2012ss} for strings in $AdS_{3}\times S^{3}\times S^{3}\times S^{1}$,
when scattering particles of different mass:
\begin{enumerate}
\item Preserve the AFS phase's form $\theta_{\text{AFS}}=ih/W\sum_{r=2}^{\infty}\left[Q_{r}Q'_{r+1}-Q_{r+1}Q'_{r}\right]$,
which gives \eqref{eq:our-AFS-phase}.
\item Start this sum from $r=1$, giving one more term, \eqref{eq:our-beta-phase}
with $\beta=1$. 
\item Add an extra string frame-like phase, as in \eqref{eq:our-S11-ansatz},
with $\Delta=-1-2\Gamma$. 
\end{enumerate}
Testing these against the tree-level near-BMN scattering \cite{Sundin:2013ypa},
we find that given the first point, the other two are obligatory.
And all parameters but $\gamma$ and $\Gamma$ are then fixed. The
crossing equations (up to one-loop order) give a similar constraint;
in particular the first point requires the second. Finally the semiclassical
limit of the Bethe equations gives another, compatible constraint
which also relates $\gamma$ and $\Gamma$.

This leaves one free parameter. We conjecture that this is $\gamma=0$,
and thus $\Gamma=-\tfrac{3}{2}$, because known string solutions can
be placed in one or both $S^{3}$ factors, and this fact must be reflected
in the Bethe equations. As $\alpha\to0,1$ we approach $AdS_{3}\times S^{3}\times T^{4}$
with a unit radius sphere, and thus should recover the usual $su(2)$
equation.%
\footnote{In particular we expect the usual AFS phase. This is the reason for
not allowing some power of $\sigma_{\text{one}}$ in $S^{11}$ \eqref{eq:our-S11-ansatz}. %
} At $\alpha=\tfrac{1}{2}$ we can place exactly the same solution
in each $S^{3}$, and the situation is very similar to that studied
in $AdS_{4}\times CP^{3}$ by \cite{Gromov:2008qe}, where it was
necessary to scale the coupling $h$ by the mass of the particles.

The S-matrix has been compared to one-loop worldsheet scattering only
for massive modes at $\alpha=1$, when the background is $AdS_{3}\times S^{3}\times T^{4}$
\cite{Sundin:2013ypa,Sundin:2014sfa,Roiban:2014cia}. This is only
sensitive to the equal-mass phase $S^{11}$, and is thus unaffected
by our proposal.%
\footnote{But note aside that both \eqref{eq:our-beta-phase}  and \eqref{eq:AFS-sum-QQ}
are zero at order $1/h^{2}$ in the BMN limit. %
} 

In the case of $AdS_{3}\times S^{3} \times T^{4}$ with mixed NS-NS and R-R flux, some issues of 
how to correctly define the AFS phase were discussed in \cite{Babichenko:2014yaa}. In that case, 
the dispersion relation is $E(p)=\sqrt{M^2+4h^2(1-\chi^{2})\sin^{2}(p/2)}$ with $M^2(p) = \left(1\pm \chi h p\right)^2$, differing 
for left and right sectors (with $\chi = 0$ for pure R-R). But no differences from the earlier proposal of \cite{Hoare:2013ida} are claimed at tree level.

The dressing phase also matters a great deal in the quantum Bethe
equations; this is of course how the one-loop phase was discovered
\cite{Beisert:2005cw,Hernandez:2006tk}. Comparisons of such results
against one-loop energy corrections to spinning strings have been
published in \cite{Beccaria:2012kb,Beccaria:2012pm}, and (unlike
$AdS_{5}\times S^{5}$) they do not yet see perfect agreement. 

\vspace{3pt}

\section*{Acknowledgements}

We thank Diego Bombardelli, Romuald Janik, Per
Sundin and especially Olof Ohlsson Sax for discussions and comments. Michael is supported by an NRF Innovation
fellowship. In\^{e}s was partially supported by the FCT--Portugal
fellowship SFRH/BPD/69696/2010 and by the NCN grant 2012/06/A/ST2/00396.
We thank CERN for hospitality.

\bibliographystyle{my-JHEP-5}
\bibliography{complete-library-processed}

\end{document}